\documentclass[aps,prd,nobibnotes,twocolumn,superscriptaddress,longbibliography]{revtex4-1}

\usepackage{amsfonts}
\usepackage{mathrsfs}
\usepackage{amsmath}
\usepackage{color}
\usepackage{graphicx}
\usepackage{bm}
\usepackage{amssymb}
\usepackage{xspace}
\usepackage{epstopdf}
\usepackage{dcolumn}
\usepackage{longtable}
\usepackage{multirow}
\usepackage[colorlinks=true, letterpaper=true, pdfstartview=FitV, linkcolor=blue, citecolor=blue, urlcolor=blue]{hyperref}

\begin{document}


\title{Two-dimensional CoSe structures: Intrinsic magnetism, strain-tunable anisotropic valleys, magnetic Weyl point, and antiferromagnetic metal state}

\author{Bo Tai}
\affiliation{Research Laboratory for Quantum Materials, Singapore University of Technology and Design, Singapore 487372, Singapore}

\author{Weikang Wu}
\affiliation{Research Laboratory for Quantum Materials, Singapore University of Technology and Design, Singapore 487372, Singapore}

\author{Xiaolong Feng}
\affiliation{Research Laboratory for Quantum Materials, Singapore University of Technology and Design, Singapore 487372, Singapore}

\author{Yalong Jiao}
\affiliation{Theoretische Chemie, Technische Universit\"{a}t Dresden, Dresden 01062, Germany}
\affiliation{Research Laboratory for Quantum Materials, Singapore University of Technology and Design, Singapore 487372, Singapore}

\author{Jianzhou Zhao}
\affiliation{Research Laboratory for Quantum Materials, Singapore University of Technology and Design, Singapore 487372, Singapore}
\affiliation{Sichuan Co-Innovation Center for New Energetic Materials,
Southwest University of Science and Technology, Mianyang 621010, China}

\author{Yunhao Lu}
\email{luyh@zju.edu.cn}
\affiliation{Zhejiang Province Key Laboratory of Quantum Technology and Device, Department of Physics, Zhejiang University, Hangzhou 310027, China}

\author{Xian-Lei Sheng}
\email{xlsheng@buaa.edu.cn}
\affiliation{Key Laboratory of Micro-nano Measurement-Manipulation and Physics (Ministry of Education), School of Physics, Beihang University, Beijing 100191, China}
\affiliation{Research Laboratory for Quantum Materials, Singapore University of Technology and Design, Singapore 487372, Singapore}

\author{Shengyuan A. Yang}
\affiliation{Research Laboratory for Quantum Materials, Singapore University of Technology and Design, Singapore 487372, Singapore}
\affiliation{Center for Quantum Transport and Thermal Energy Science, School of Physics and Technology, Nanjing Normal University, Nanjing 210023, China}


\begin{abstract}
The interplay between magnetism, band topology, and electronic correlation in low dimensions has been a fascinating subject of research. Here, we propose two-dimensional (2D) material systems which demonstrate such an interesting interplay. Based on first-principles calculations and structural search algorithms, we identify three lowest energy 2D CoSe structures, termed as the $\alpha$-, $\beta$-, and $\gamma$-CoSe. {We show that $\alpha$- and $\beta$-CoSe are two rare examples of 2D antiferromagnetic metals, which are related to their Fermi surfaces nesting features, and meanwhile, $\gamma$-CoSe is a ferromagnetic metal. They possess a range of interesting physical properties, including anisotropic valleys connected by crystalline symmetries, strain-tunable valley polarization, strain-induced metal-semiconductor and/or magnetic phase transitions, as well as topological band features such as the magnetic Weyl point and the magnetic Weyl loop. Remarkably, all the topological features here are robust against spin-orbit coupling.} Some experimental aspects of our predictions have been discussed.
\end{abstract}

\maketitle


\section{\label{sec:level1}INTRODUCTION}
Two-dimensional (2D) materials has been attracting tremendous research interest in recent years~\cite{butler2013progress,tan2017recent,manzeli20172d,low2017polaritons}. With a thickness of only one or a few atomic layers, 2D materials naturally enjoy the advantages of large surface-to-volume ratio, excellent mechanical flexibility, and easy tunability via applied fields or chemical functionalization. Besides, the intricate interplay between quantum effects and reduced dimensionality may give rise to  a range of surprising physics, especially for 2D materials with transition metal elements. For example, it has been reported that the superconductivity in layered FeSe can be remarkably enhanced when approaching the 2D limit and placed on a substrate. Its transition temperature can be increased from 8 K for the bulk to 65-109 K for a single layer on SrTiO$_3$~\cite{Wang2012,he2013phase,tan2013interface,peng2014tuning,miyata2015high,ge2015superconductivity,zhou2018antiferromagnetic}. The recent discovery of intrinsic magnetism in 2D layers of CrI$_3$~\cite{huang2017layer,huang2018electrical}, Cr$_2$Ge$_2$Te$_6$~\cite{gong2017discovery}, and Fe$_3$GeTe$_2$~\cite{deng2018gate} is another example. The magnetic ordering from the transition metal $d$ orbitals is surprisingly robust, with Curie temperatures above the liquid helium or even liquid nitrogen temperature. These fascinating findings have triggered a surge of efforts to explore novel 2D transition metal compounds.

In 2016, Zhou~\emph{et al.}~\cite{zhou2016metastable} synthesized a new transition metal compound, the tetragonal CoSe, by using a topochemical deintercalation approach. The bulk material has a layered structure, and is isostructural to the famous FeSe superconductor. Although superconductivity has not been detected, measurements have shown that bulk CoSe is a ferromagnetic metal with in-plane magnetic moments~\cite{zhou2016metastable,wilfong2018frustrated}. Because of its layered structure, it is possible to thin down the material towards the 2D regime. Indeed, Ma~\emph{et al.}~\cite{ma2019phase} have fabricated ultrathin CoSe nanoplates with a thickness down to 2.3 nm (about 4-5 layers), using a chemical vapor deposition method, and revealed its interesting thickness-dependent transport property. Shen \emph{et al.}~\cite{shen2018evolution} and Liu \emph{et al.}~\cite{liu2018anti} fabricated ultrathin films of tetragonal CoSe down to monolayer limit on SrTiO$_3$ substrate via molecular beam epitaxy approach, targeting at electronic features that might be connected to superconductivity. Very recently, the application of CoSe thin films in infrared photodetectors was also investigated~\cite{liang2020ferromagnetic}. Despite these exciting progress, there are important questions awaiting to be answered: What are the stable or metastable structural phases for CoSe in the 2D limit? And what are the special physical properties of these 2D CoSe phases?

\begin{figure}
	\includegraphics[width=0.5\textwidth]{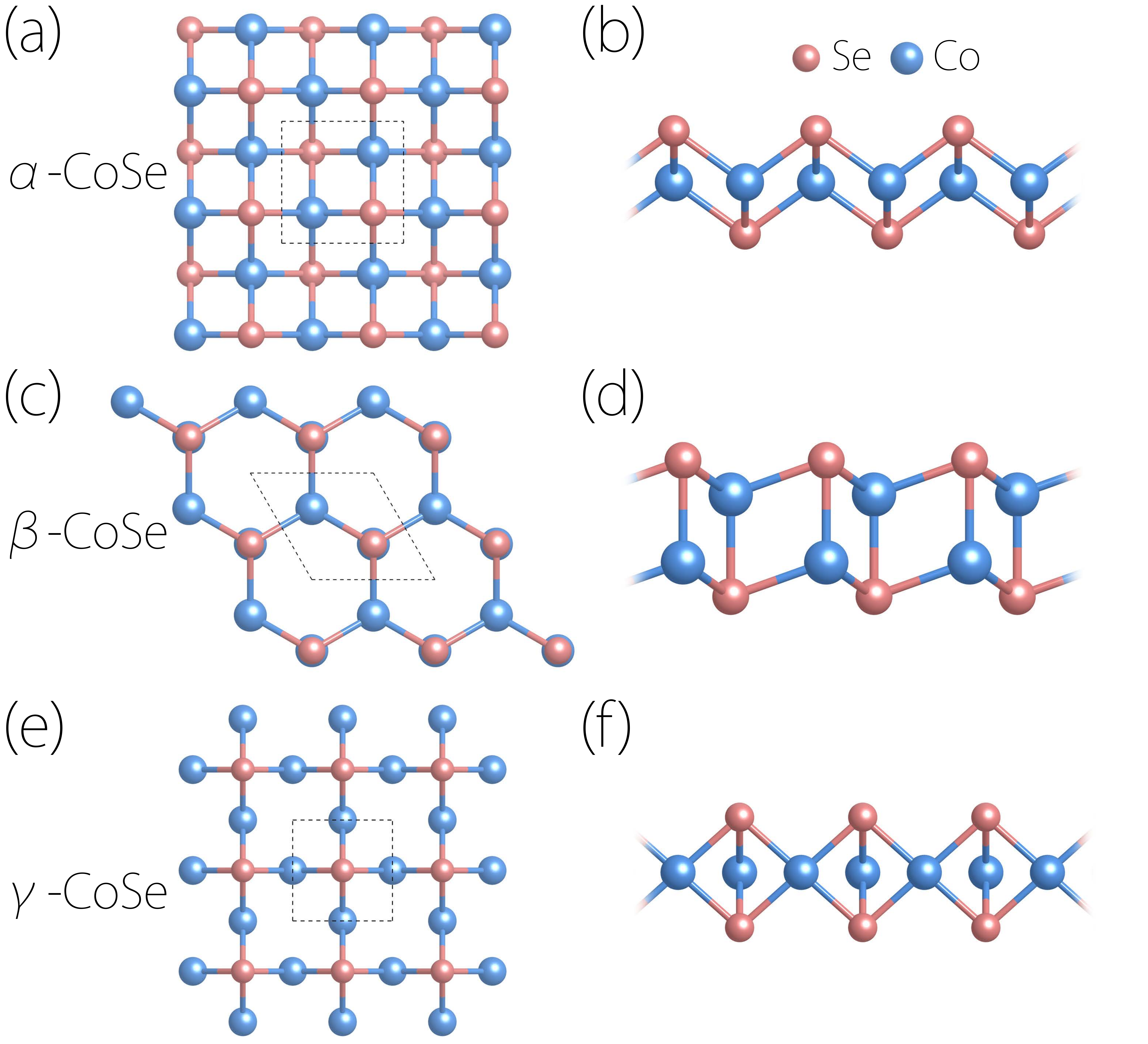}
	\caption{Top and side views of the atomic structures of (a,b) $\alpha$-CoSe, (c,d) $\beta$-CoSe, and (e,f) $\gamma$-CoSe.}
	\label{fig_structures}
\end{figure}

Motivated by these questions and the experimental advances mentioned above, in this work, we theoretically explore the possible 2D CoSe structures in the monolayer limit and investigate their electronic properties. By carrying out comprehensive structural search combined with first-principles calculations, we identify the single layer of the tetragonal CoSe phase (termed as $\alpha$-CoSe) as the global minimum in the 2D monolayer limit. Meanwhile, we also predict another two metastable 2D CoSe structural phases (termed as $\beta$ and $\gamma$ phases). These structures are illustrated in Fig.~\ref{fig_structures}. We show that these monolayer structures have good dynamical and thermal stabilities. Interestingly, the different crystal structures result in different magnetic ground states: {Both $\alpha$- and $\beta$-CoSe show out-of-plane antiferromagnetism (AFM), whereas $\gamma$-CoSe shows in-plane ferromagnetism (FM).} From Monte-Carlo simulations, we find that their magnetic orderings are fairly robust, with estimated transition temperatures above the liquid nitrogen temperature. More importantly, we discover in these 2D materials several unusual features.
{$\alpha$-CoSe hosts a pair of anisotropic valleys connected by the crystalline symmetry rather than the time reversal symmetry as in the MoS$_2$-family materials. Hence, the valley degree of freedom can be readily controlled by lattice strain, and strain can drive a metal-semiconductor phase transition in $\alpha$-CoSe. For both 
$\alpha$- and $\beta$-CoSe, we find the AFM coexists with a metallic ground state, which is rather unusual and is associated with the Fermi surface nesting character in their band structures.}
$\gamma$-CoSe is a ferromagnetic metal. It features a pair of 2D magnetic Weyl points below the Fermi level, protected by a twofold rotational symmetry. And when the magnetization is rotated to the $z$ direction, the two Weyl points will evolve into a single Weyl loop, which is 100\% spin polarized. Importantly, both the magnetic Weyl points and the magnetic Weyl loop found here are robust against SOC. Our work unveils multiple 2D structural phases for a new transition metal compound. Their rich physical properties may lead to promising spintronic and electronic applications.

\section{\label{sec:level1}METHOD}

The 2D crystal structural search was performed by using the evolutionary algorithm implemented in the USPEX code~\cite{oganov2006crystal, lyakhov2013new, oganov2011evolutionary} combined with the density functional theory (DFT) calculations using the Vienna \emph{ab-initio} simulation package (VASP)~\cite{kresse1993ab, kresse1996efficient}. In the search, the number of atoms in a unit cell was limited to 12 (i.e., six formula units), and the thickness of the 2D layer was limited to 4~\AA. The single layer tetragonal CoSe structure ($\alpha$-CoSe) was used as a seed in the first generation. New structures were generated by carefully designed variation operators, such as heredity and soft mutation, and were fully relaxed. The relaxed energy was used for selecting structures as parents for the next generation of structures. The whole search evolved 50 generations with 40 structures in each generation.

In our DFT calculation, the projector augmented wave method~\cite{blochl1994projector} was used to describe the eletron-ion interactions. The generalized gradient approximation with the Perdew-Burke-Ernzerhof (PBE) realization~\cite{perdew1996generalized} was adopted for the exchange correlation functional. To account for the important correlation effect associated with the Co-$3d$ orbitals, we included the Hubbard $U$ correction via the PBE$+U$ method~\cite{dudarev1998electron}. Here, the $U$ value was taken to be 4 eV according to {Ref.~\cite{dalverny2010interplay}}. A vacuum layer of 15 \AA\ was added to avoid artificial interactions between periodic images. The plane wave energy cutoff was set to be 600 eV, and the BZ was sampled with $\Gamma$-centered $k$ mesh with size of $15 \times 15\times 1$. The structures were fully optimized with the energy and force convergence criteria of 10$^{-7}$ eV and 10$^{-4}$ eV/\AA, respectively. Our band structure results have also been verified by using the hybrid functional (HSE06) approach~\cite{heyd2003hybrid}.

The phonon spectra of the materials were calculated using the PHONOPY code through the DFPT approach~\cite{togo2015first} on a $4 \times 4 \times 1$ supercell. The thermal stability was checked by performing the \emph{ab-initio} molecular dynamics (AIMD) simulations on a $5 \times 5 \times 1$ supercell. In the simulation, the NVT canonical sampling was performed by integrating the equations of motion at 2 fs time intervals, and the temperature was controlled via a Nos\'{e}-Hoover thermostat. At each time step, the total energy was evaluated to an accuracy of 10$^{-4}$ eV/cell with a plane-wave energy cutoff of 300 eV.

\begin{figure}
	\includegraphics[width=0.5\textwidth]{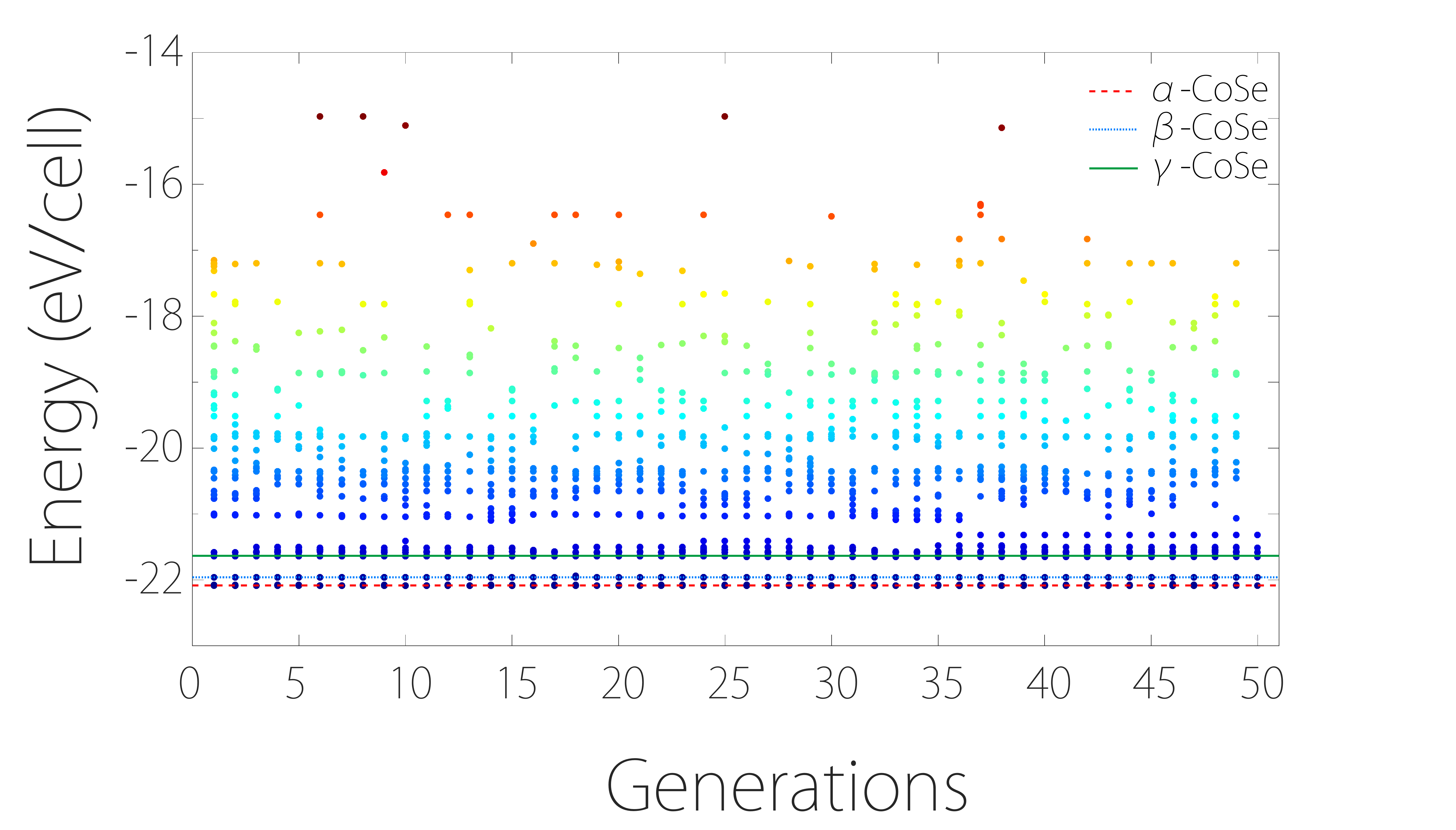}
	\caption{Overview of the structural search. The structures are ordered by their energies. The three low energy structures are highlighted.}
	\label{fig_GA}
\end{figure}

\begin{figure}
	\includegraphics[width=0.5\textwidth]{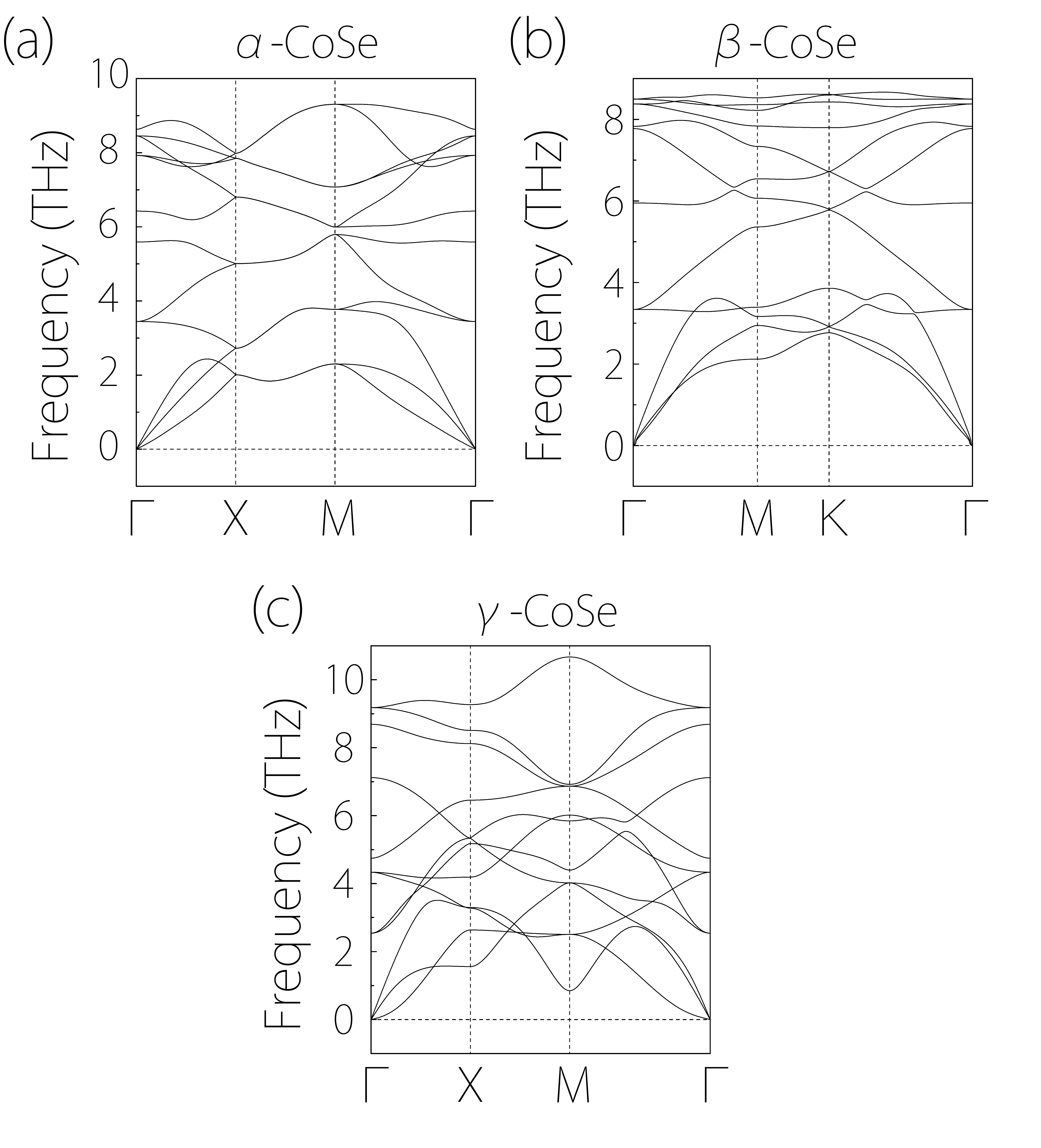}
	\caption{Calculated phonon spectra for (a) $\alpha$-CoSe, (b) $\beta$-CoSe, and (c) $\gamma$-CoSe. }
	\label{fig_phonon}
\end{figure}

\section{\label{sec:level1}2D Cobalt Selenide structures}

We have performed comprehensive global minimum structural search for 2D CoSe structures using the evolutionary algorithm, as described in the last section. Each generated structure has been fully relaxed using VASP, and the structures have been sorted by energy. Figure~\ref{fig_GA} gives an overview of our structural search.

We find that the lowest-energy structure [$\alpha$-CoSe, see Fig.~\ref{fig_structures}(a,b)] is isostructural to the single layer of the tetragonal CoSe synthesized in the previous experiments~\cite{zhou2016metastable,wilfong2018frustrated,shen2018evolution,liu2018anti,ma2019phase}. Besides, there are two well defined gaps above this global minimum (see Fig.~\ref{fig_GA}), suggesting that the two structures ($\beta$- and $\gamma$-CoSe, see Fig.~\ref{fig_structures}) correspond to meta-stable local minima. In the following, we will focus on these three low energy structures. 
 (The next low-energy structure, termed as $\delta$-CoSe, is also found to be meta-stable. See the Supplemental Material~\cite{OtherPair}.)

The three 2D CoSe structures are illustrated in Fig.~\ref{fig_structures}. Both $\alpha$- and $\gamma$-CoSe have a square type lattice, whereas $\beta$-CoSe has a hexagonal lattice. $\alpha$-CoSe is isostructural to the single layer FeSe. Each Co atom is surrounded by four Se atoms, forming a tetrahedral coordination. In comparison, for $\gamma$-CoSe, the top view of the lattice [Fig.~\ref{fig_structures}(e)] resembles a Lieb lattice. While the Co atoms all lie in the same atomic plane, each Se site has two atoms lying on top of each other, off the Co plane. As for $\beta$-CoSe, its top view shows a bipartite honeycomb lattice. However, each site in this honeycomb is actually occupied by a Co-Se pair lying on top of each other [see Fig.~\ref{fig_structures}(c,d)]. We note that for $\alpha$- and $\gamma$-CoSe, the Co-Co distance (2.307 \AA\ for $\alpha$, 2.358 \AA\ for $\gamma$) is less than that in the Co metal (2.506 \AA), whereas for $\beta$-CoSe, this distance (2.607 \AA) is slightly larger.
The detailed structural parameters are presented in Table~I. {(In the table, the enthalpy of
formation is with reference to the bulk tetragonal CoSe, defined as $E=E_\text{2D}-E_\text{bulk}$, where $E_\text{2D}$ is the total enthalpy for the 2D structure, and $E_\text{bulk}$ is the enthalpy \emph{per layer} for the bulk structure.)} The structural data files are also provided in the Supplemental Material~\cite{OtherPair}. 

To assess the stability of these 2D structures, we have calculated their phonon spectra. As shown in Fig.~\ref{fig_phonon}, the phonon spectra exhibit no soft mode, indicating that the three structures are dynamically stable.
The thermal stability is investigated by the AIMD simulations. We have performed the simulation for each of the three CoSe structures at 300 K. The result confirms that all the three lattice structures are well maintained against the thermal fluctuations at room temperature (see the Supplemental Material~\cite{OtherPair}).


\begin{table*}[!htp]
\centering
\caption{Structural and magnetic properties of the three 2D CoSe structures. Here, the magnetic moment is the one on each Co site. The moment on the Se site is negligible. }
\renewcommand{\arraystretch}{1.6}
\begin{tabular}{p{1.9cm}<{\centering}p{1.9cm}<{\centering}p{1.9cm}<{\centering}p{1.9cm}<{\centering}p{1.9cm}<{\centering}p{1.9cm}<{\centering}p{1.9cm}<{\centering}p{1.9cm}<{\centering}p{1.9cm}<{\centering}}\hline\hline
      & Lattice & Layer group & Space group & Lattice constant & Magnetism & Magnetic moment  & Easy axis & Enthalpy \\ \hline
$\alpha$-CoSe & Square          & 64      & P4/nmm      & 3.74 \AA          & AFM       & 2.1$\mu_B$      & $\langle 001\rangle$ & 0.05 eV\\
$\beta$-CoSe  & Hexagonal       & 72      & P$\bar{3}$m1    & 3.74 \AA          & AFM       & 2.2$\mu_B$      & $\langle 001\rangle$ &0.21 eV\\
$\gamma$-CoSe & Square          & 61      & P4/mmm      & 3.51 \AA          & FM        & 1.9$\mu_B$      & $\langle 100\rangle$ & 0.51 eV\\ \hline\hline
\end{tabular}
\end{table*}

\begin{figure}
	\includegraphics[width=0.3\textwidth]{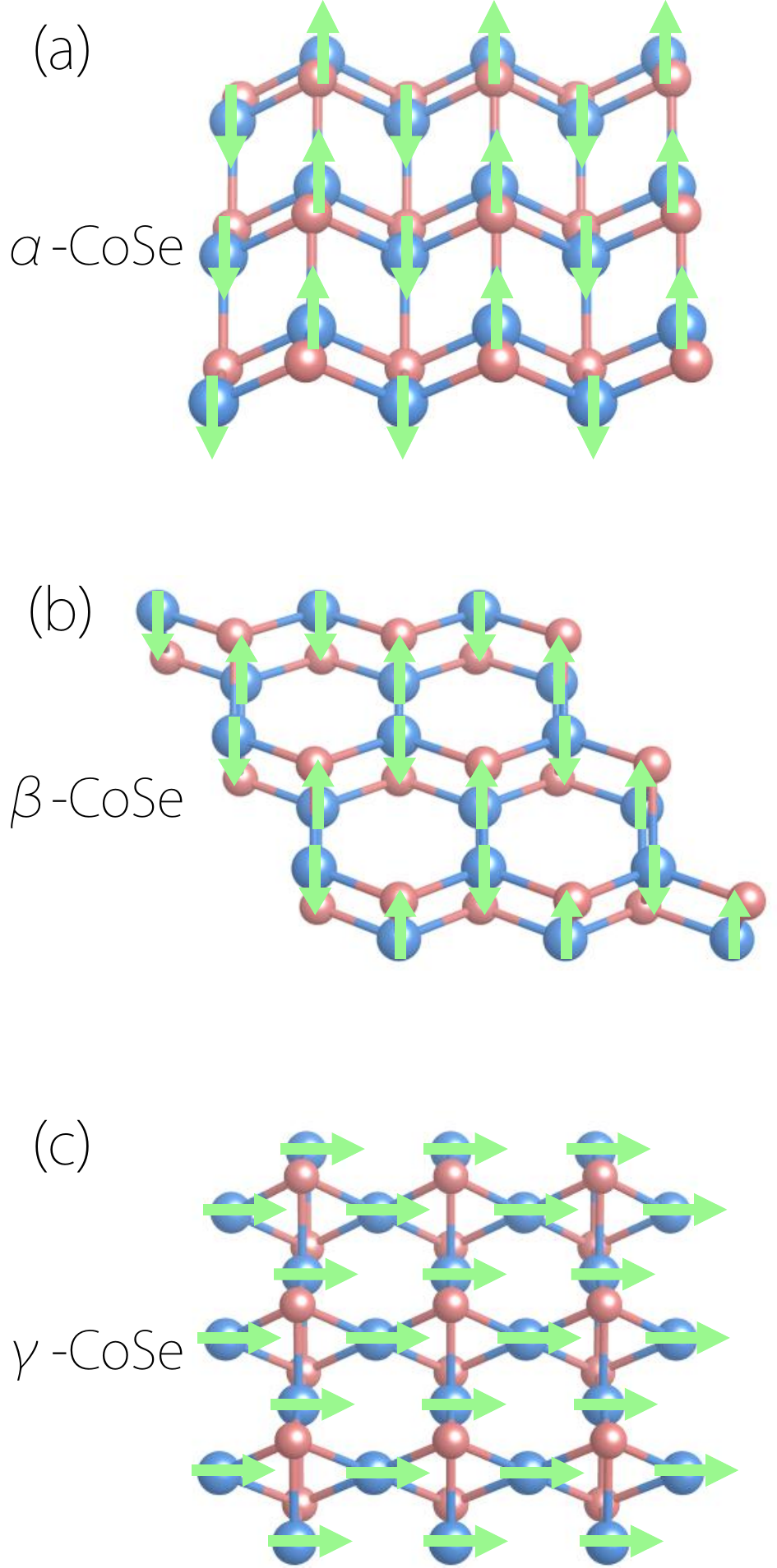}
	\caption{Illustration of the ground state magnetic configurations for (a) $\alpha$-CoSe, (b) $\beta$-CoSe, and (c) $\gamma$-CoSe. Here, we are taking a perspective view. For $\alpha$- and $\beta$-CoSe, the magnetic moments are along the out-of-plane direction, whereas for $\gamma$-CoSe, they are along the in-plane $x$ direction. }
	\label{fig_mag}
\end{figure}

\section{Magnetic property}

Co is a $3d$ transition metal element. Materials containing Co often exhibit magnetic orderings in the ground state. Hence, we  need to first pin down the ground-state magnetic configuration for each of the three CoSe 2D structures.

Based on first-principles calculations, we have compared the energies of the different types of magnetic configurations, including the nonmagnetic (NM), the FM, and several possible AFM configurations. In the process, we have included the SOC to determine the magnetic anisotropy. The results are illustrated in Fig.~\ref{fig_mag} and presented in Table~I.
{The calculation results indicate that both $\alpha$- and $\beta$-CoSe have AFM ground state; whereas $\gamma$-CoSe is FM. The magnetic moments for $\alpha$- and $\beta$-CoSe prefer the out-of-plane ($z$) direction; whereas for $\gamma$-CoSe, the preferred direction is the in-plane $x$ direction.} For all three structures, the magnetic moments are mainly distributed on the Co ions, with a magnitude $\sim 2 \mu_B$.

We have estimated the magnetic transition temperatures for these states by performing the Monte-Carlo simulations based on a classical spin model~\cite{evans2014atomistic}:
\begin{equation}
  H=-\sum_{\langle i,j\rangle}J_{ij}\bm S^i\cdot \bm S^j -K\sum_i (S^i_\alpha)^2.
\end{equation}
Here, the spin vectors are normalized, $i$ and $j$ label the Co sites, the first term represents the exchange coupling between the nearest neighbors $i$ and $j$, the second term represents the magnetic anisotropy, and $\alpha$ refers to the easy axis direction. The values of the parameters $J_{ij}$ and $K$ are determined from the DFT calculation (see \cite{OtherPair} for details). From the Monte-Carlo simulations, the estimated N\'{e}el temperatures for $\alpha$- and $\beta$-CoSe are around {300 K} and {195 K}, respectively; while the Curie temperature for $\gamma$-CoSe is about {350 K}. These values are above the liquid nitrogen temperature, indicating that the magnetism in these 2D CoSe structures are fairly robust.

\section{Electronic property}
After fixing the magnetic ground states, in the following, we shall turn to the electronic band structure properties of these 2D CoSe structures.

\begin{figure}
	\includegraphics[width=0.4\textwidth]{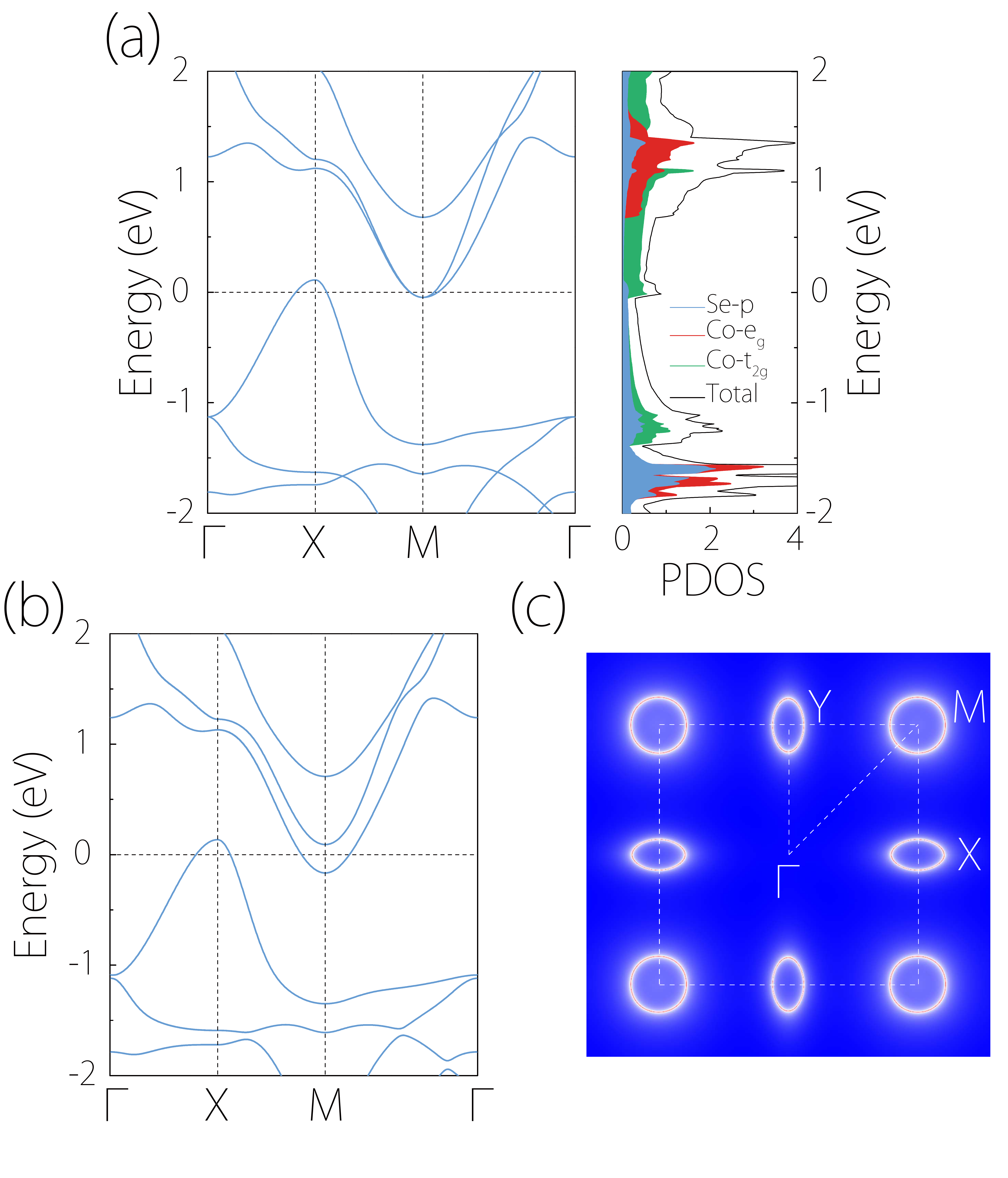}
	\caption{(a) Calculated band structure and PDOS for $\alpha$-CoSe in the absence of SOC. (b) Band structure of $\alpha$-CoSe with SOC included. (c) Fermi surface for $\alpha$-CoSe (SOC included). There are two anisotropic hole valleys at $X$ and $Y$ points. }
	\label{fig_alpha-band}
\end{figure}

\begin{figure}
	\includegraphics[width=0.45\textwidth]{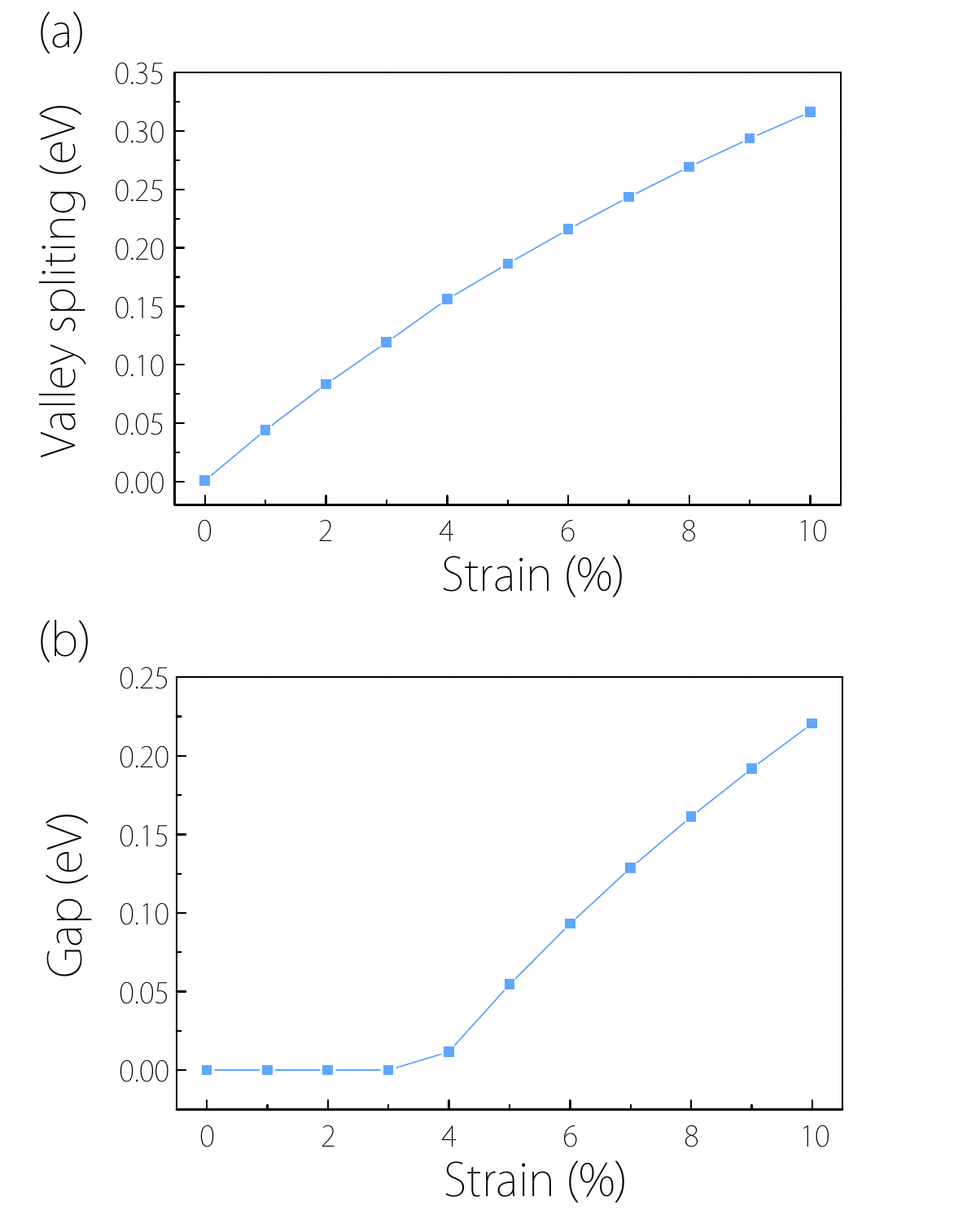}
	\caption{(a) Valley splitting and (b) band gap of $\alpha$-CoSe versus applied uniaxial strain along the $x$ direction.}
	\label{fig_alpha-strain}
\end{figure}

\begin{figure}
	\includegraphics[width=0.46\textwidth]{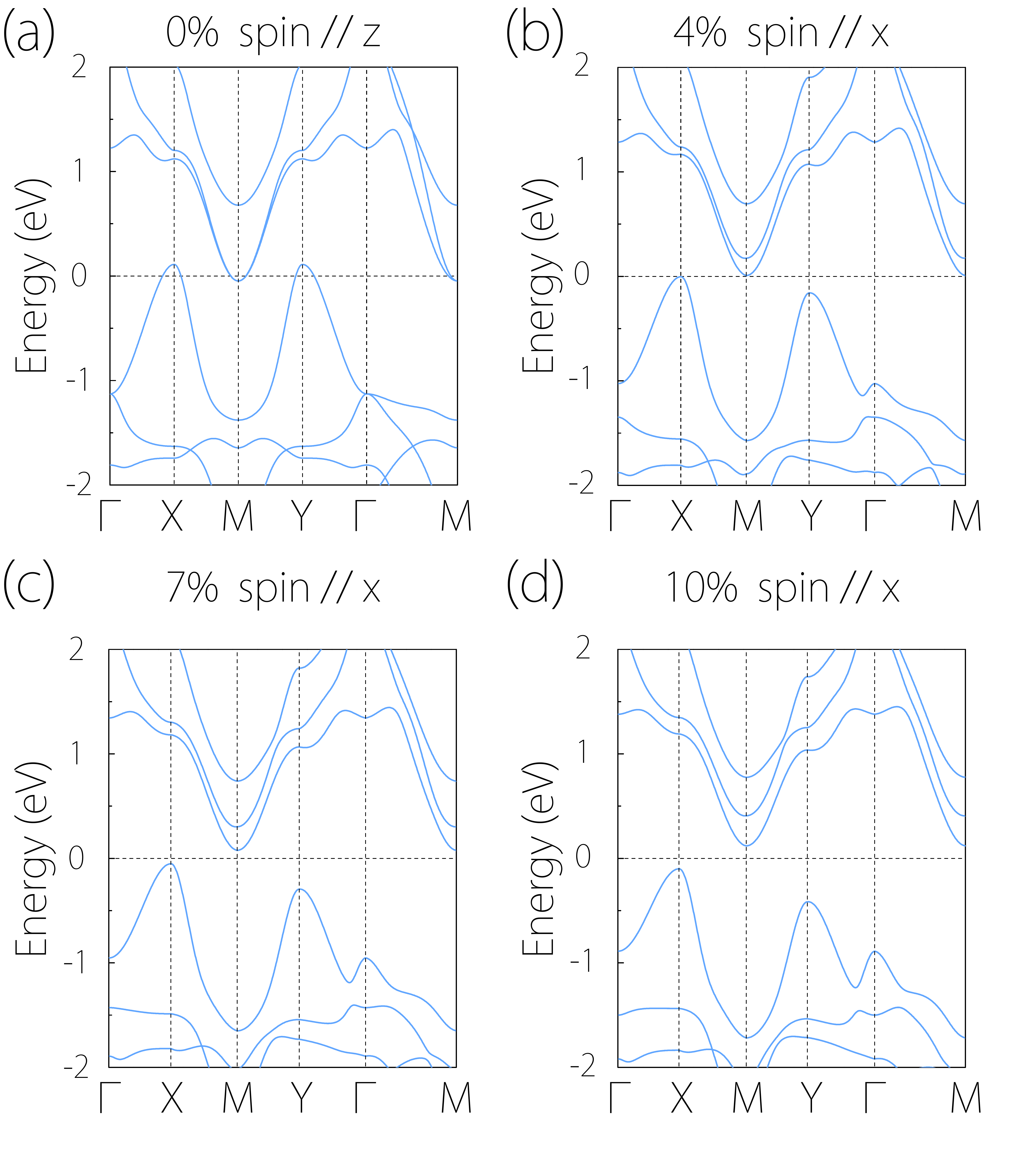}
	\caption{Calculated band structures of $\alpha$-CoSe (SOC included) with several applied uniaxial strains $\varepsilon_x$ along the $x$ direction: (a) $\varepsilon_x$=0$\%$, (b) $\varepsilon_x$=4$\%$, (c) $\varepsilon_x$=7$\%$ and (d) $\varepsilon_x$=10$\%$.}
	\label{fig_alpha-strainband}
\end{figure}

{
\subsection{$\alpha$-CoSe: Strain tunable anisotropic valleys}
Let us first consider $\alpha$-CoSe. Its ground state has AFM with magnetization along the out-of-plane ($z$) direction. Figure~\ref{fig_alpha-band}(a \& b) shows the calculated band structures for this ground state without and with SOC, along with the orbital projected density of states (PDOS). One observes that $\alpha$-CoSe is an AFM metal.
This is a very interesting observation, because the AFM ordering is usually accompanied by an insulating state. AFM metals are quite rare, not to mention that our current example is in 2D. Interestingly, $\beta$-CoSe is also found to be an AFM metal. In Sec.~\ref{beta}, we will discuss this feature and show that the AFM metal states in these two CoSe structures are associated with
spin density waves arising from certain Fermi surface nesting features.

From Fig.~\ref{fig_alpha-band}(b), one observes that the conduction and valence bands have an indirect overlap in energy, which results in one electron pocket around $M$ and two hole pockets at $X$ and $Y$ points [see Fig.~\ref{fig_alpha-band}(c)]. The orbital analysis shows that  the electron pocket is mostly contributed by the $p$ orbitals of Se, whereas 
the hole pockets are mostly from the $t_{2g}$ orbitals of Co. Notably, the two inequivalent hole pockets endow the hole carriers in $\alpha$-CoSe with a valley degree of freedom. And the valleys here exhibit two important features distinct from the well studied ones in 2D MoS$_2$-family materials. First, the two $X$ and $Y$ valleys here are connected by the 
crystalline symmetry $\{c_{4z}|\frac{1}{2}00\}$ rather than the time reversal symmetry $\mathcal{T}$. This means the valleys can split by breaking the crystal symmetry, e.g., by lattice strain, which cannot be achieved in 2D MoS$_2$-family materials as the valleys there are connected by $\mathcal{T}$. Second, at each valley, the highest rotational symmetry is only twofold (whereas for MoS$_2$-family materials, each valley has a threefold axis). It follows that each valley should exhibit an in-plane anisotropy. For instance, at the $X$ valley, the hole effective mass is $0.74 m_e$ long the $x$ direction and $0.26m_e$ along the $y$ direction ($m_e$ is the free electron mass), differed by almost three times.

In the equilibrium state, the two valleys are degenerate due to the $\{c_{4z}|\frac{1}{2}00\}$ symmetry, and the overall physical properties are isotropic in the 2D plane. The valley degeneracy can be lifted an in-plane uniaxial strain, which breaks the fourfold rotational symmetry. In Fig.~\ref{fig_alpha-strain}(a), we plot the valley splitting as a function of the applied uniaxial strain, which exhibits an almost linear dependence up to 10\% strain. The size of the splitting can reach $\sim$0.1 eV at 2\% strain, which is quite large. In Fig.~\ref{fig_alpha-strainband}, we plot the band structures for a few representative strains. The splitting of the two valleys can be clearly observed. In addition, there are another two important features. First, there is a metal-semiconductor phase transition at about 3\% strain, as shown in Fig.~\ref{fig_alpha-strain}(b). Beyond this critical value, $\alpha$-CoSe becomes an indirect-gap semiconductor. Hence, with a small level of hole doping, we can make the holes selectively occupy a single valley by controlling the applied strains. In other words, the valley polarization and all other valley-dependent properties can be manipulated solely by strain, which is not achievable in MoS$_2$-family materials. Second, we find that around the metal-semiconductor phase transition, the magnetic anisotropy also changes: The preferred spin direction changes from the out-of-plane $z$ direction to the in-plane $x$ direction (i.e., along the tensile strain direction).

The results here indicate that $\alpha$-CoSe could serve as a new type of valleytronic platform, where the valleys are related by crystalline symmetries but not $\mathcal{T}$. This is similar to the idea proposed in Ref.~\cite{yu2020valley}, but here the realization is in a magnetic system which explicitly breaks the $\mathcal{T}$ symmetry. Besides the direct strain control of valley, we note that due to the anisotropy of each valley, the valley polarization can manifest as enhanced in-plane anisotropy in measurable physical properties. For example, the valley polarization can be detected in electric transport, through the in-plane anisotropy in the resistivity tensor elements.

}

\begin{figure}
	\includegraphics[width=0.42\textwidth]{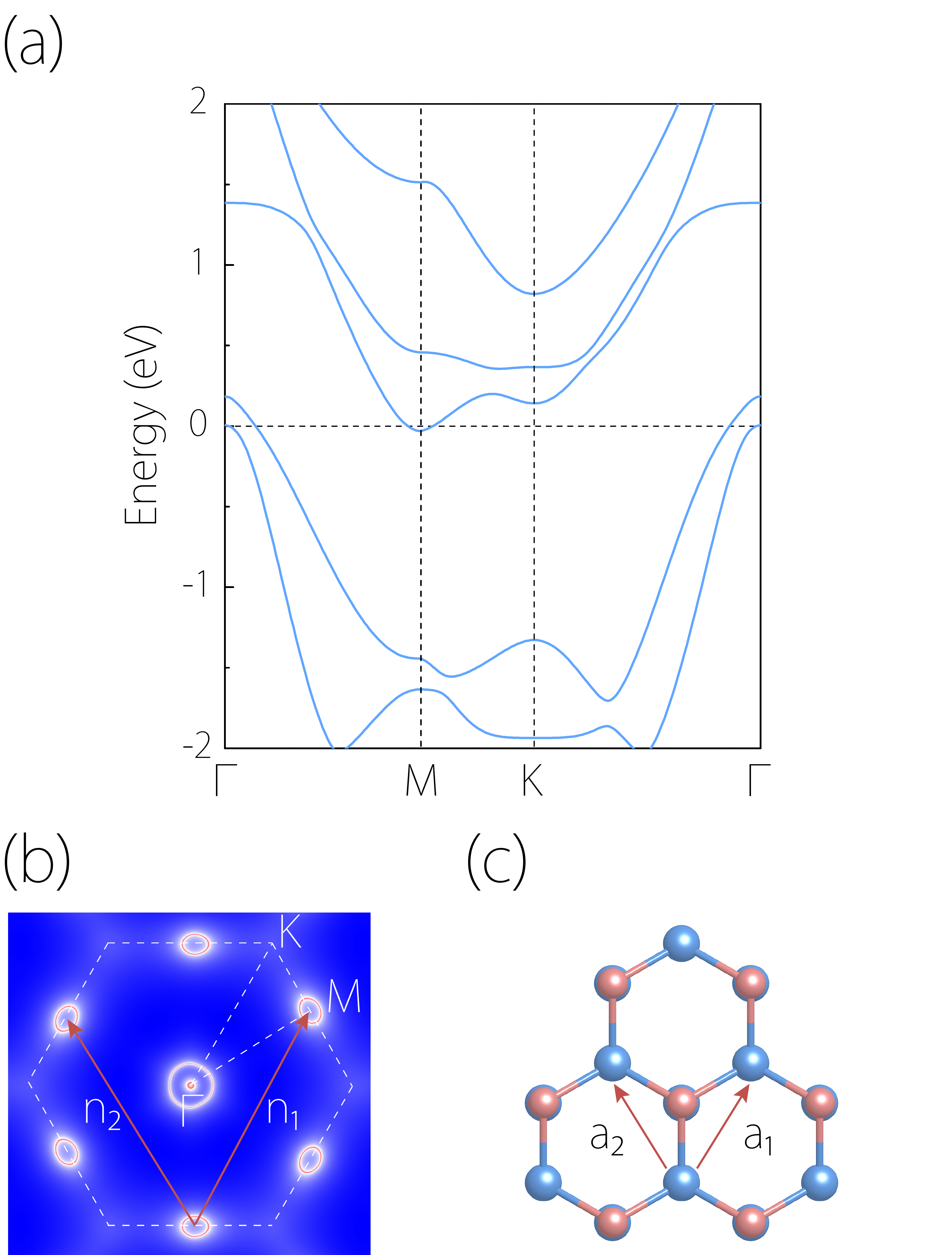}
	\caption{(a) Band structure of $\beta$-CoSe with SOC included. (b) Fermi surface of $\beta$-CoSe. $\bm n_1$ and $\bm n_2$ are the two nesting vectors. (c) Structure of $\beta$-CoSe, with the lattice vectors $\bm a_1$ and $\bm a_2$ labeled. }
	\label{fig_beta}
\end{figure}

\begin{figure}
	\includegraphics[width=0.5\textwidth]{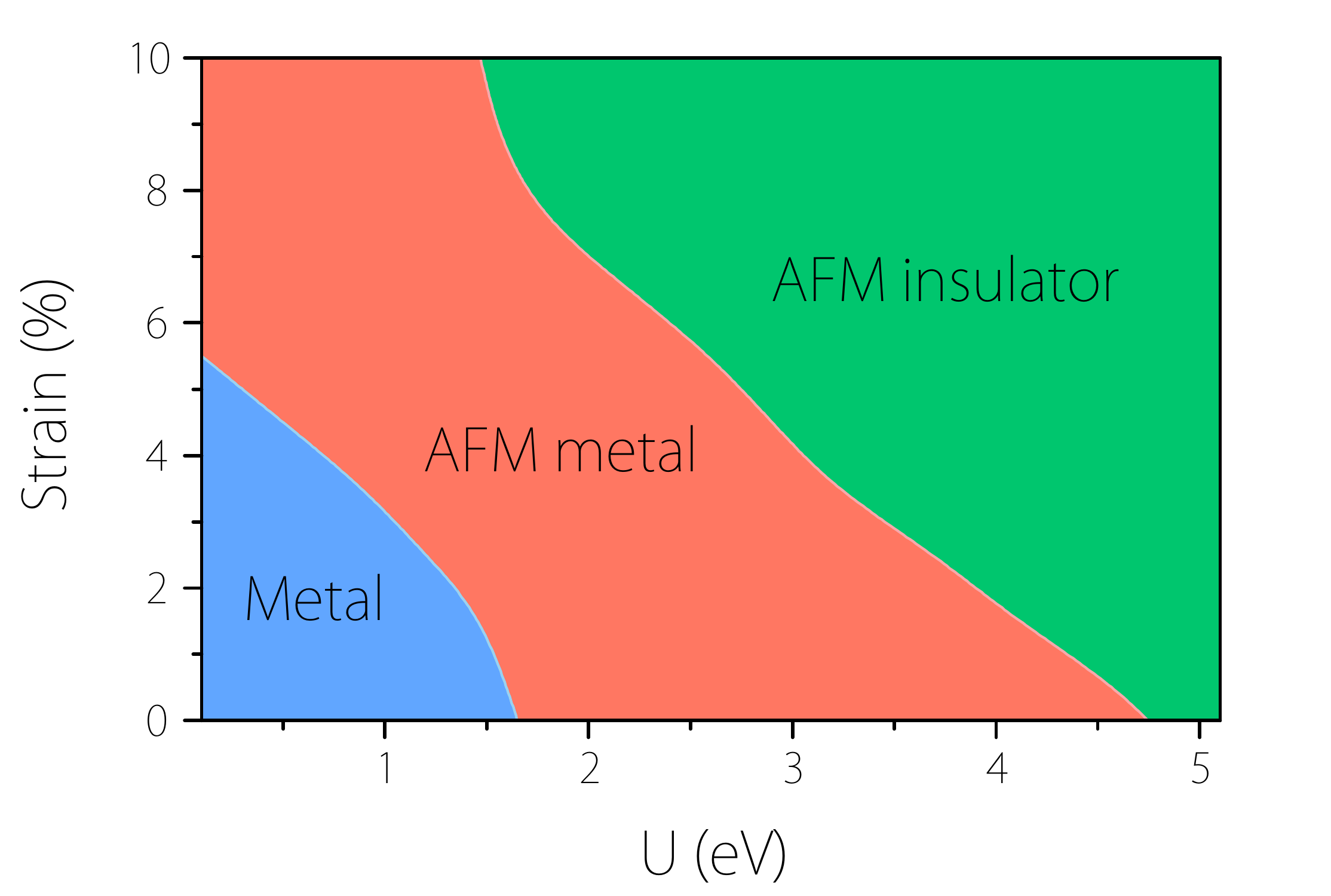}
	\caption{Phase diagram of $\beta$-CoSe with respect to $U$ and the biaxial lattice strain.}
	\label{fig_phase}
\end{figure}

\subsection{$\beta$-CoSe: AFM metal state}\label{beta}

As we have analyzed in Sec.~IV, $\beta$-CoSe has AFM in the ground state with the magnetic configuration shown in Fig.~\ref{fig_mag}(b). Figure~\ref{fig_beta}(a) shows its calculated band structure (SOC included). One notes that the system is also an AFM metal. The band gap closes indirectly due to the overlap between the conduction band minimum at $M$ and the valence band maximum at $\Gamma$. Thus, the ground state of $\beta$-CoSe is an AFM metal.
{As mentioned in the last section for $\alpha$-CoSe, AFM metals are quite unusual. In many cases, AFM metals are 
 associated with spin density waves connected with certain Fermi surface nesting features~\cite{khomskii2014transition}.}
Indeed, the wavelength for the AFM ordering [as shown in Fig.~\ref{fig_beta}(c)] is determined by the lattice vectors $\bm a_1$ and $\bm a_2$, with a magnitude of $a=|\bm a_1|=|\bm a_2|$. In Fig.~\ref{fig_beta}(b), we plot the Fermi surface of $\beta$-CoSe. It has a hole pocket at $\Gamma$ and three electron pockets at the three $M$ points. One observes the nesting feature among the three electron pockets. The nesting vectors are given by $\bm n_1$ and $\bm n_2$. One immediately notices that $\bm n_i\|\bm a_i$ ($i=1,2$), and $n=2\pi/a$, where $n$ is the magnitude of the nesting vector. This suggests that the AFM ordering in $\beta$-CoSe is closely connected with its Fermi surface nesting feature. Here, it is likely that the spin density wave is not strong enough to gap the whole Fermi surface (one important factor is the existence of itinerant carriers from the valence band maximum), so that the system remains metallic. The similar analysis also applies to $\alpha$-CoSe, where the AFM order coincides with the nesting vectors between the hole pockets. These cases are similar to well known example of the AFM metal state in Cr~\cite{khomskii2014transition}.

We note that in Fig.~\ref{fig_beta}(a), the lowest conduction band and the highest valence band only have a small overlap in energy. This suggests that the state could be quite sensitive to perturbations. Here, we consider two system parameters. One is the electron correlation parameter $U$ (here for the Co-$3d$ orbitals, the default value is 4 eV), and the other is the lattice strain. The calculated phase diagram is shown in Fig.~\ref{fig_phase}. Three phases appear in the diagram: the AFM metal, the nonmagnetic metal, and the AFM insulator. With increasing $U$, the system transforms from nonmagnetic metal to AFM metal and finally to AFM insulator, reflecting the increasing importance of electron correlation effects. Under zero strain, the transition from an AFM metal to an AFM insulator occurs at about $U=4.8$ eV. Increasing the lattice strain shows the similar trend, because the strain suppresses the electron kinetic energy, hence effectively enhancing the electron correlation effects. These behaviors are consistent with the typical picture based on the Hubbard model~\cite{khomskii2014transition}. The representative band structures for the nonmagnetic metal and AFM insulator phases can be found in the Supplemental Material~\cite{OtherPair}.

\begin{figure}
	\includegraphics[width=0.42\textwidth]{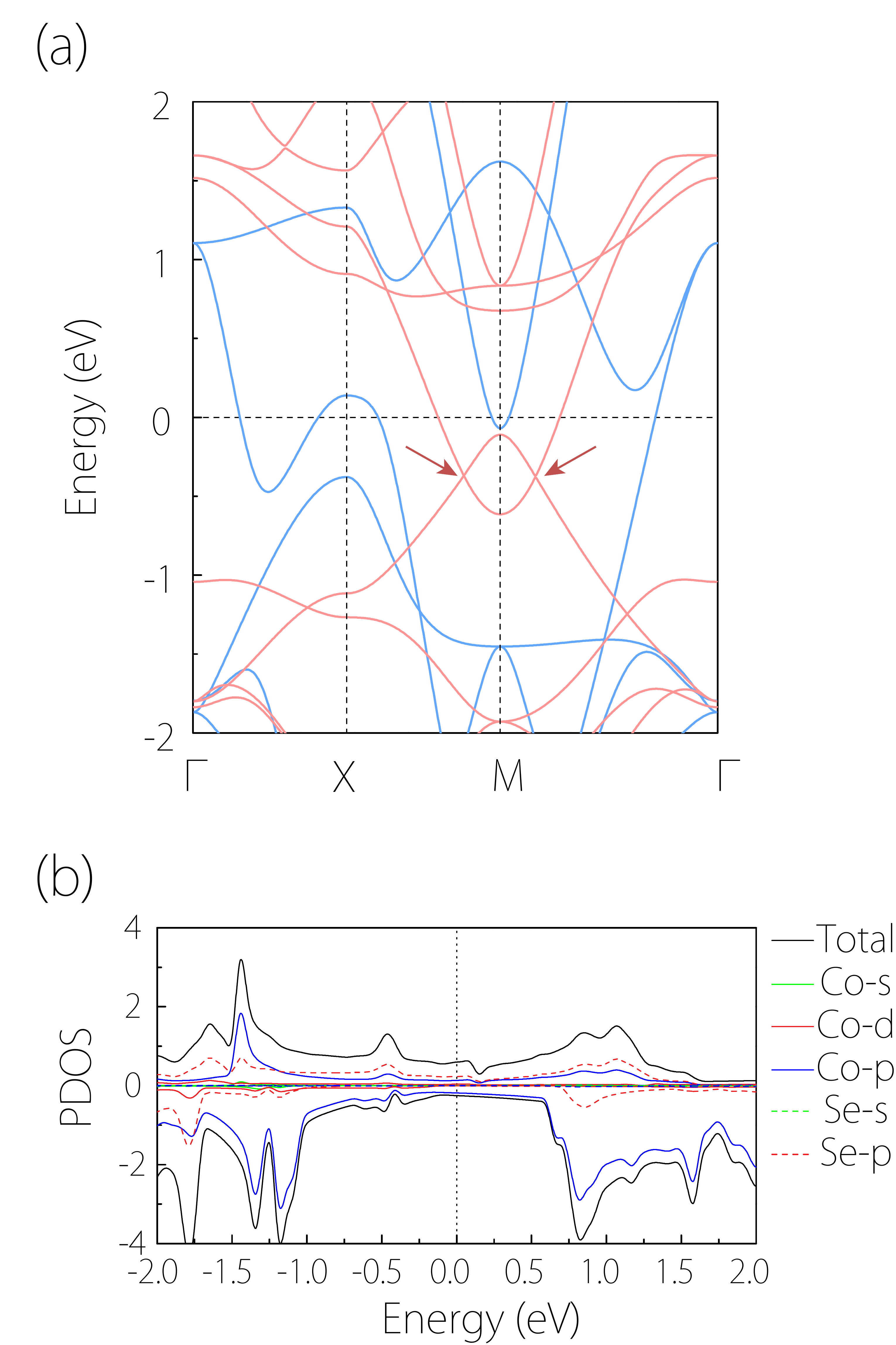}
	\caption{(a) Band structure and (b) PDOS of $\gamma$-CoSe in the absence of SOC. In (a), the two spin channels are marked by the red and the blue colors, respectively. The red arrows in (a) indicate the magnetic Weyl loop centered around $M$.}
	\label{fig_gamma-band}
\end{figure}

\begin{figure}
	\includegraphics[width=0.5\textwidth]{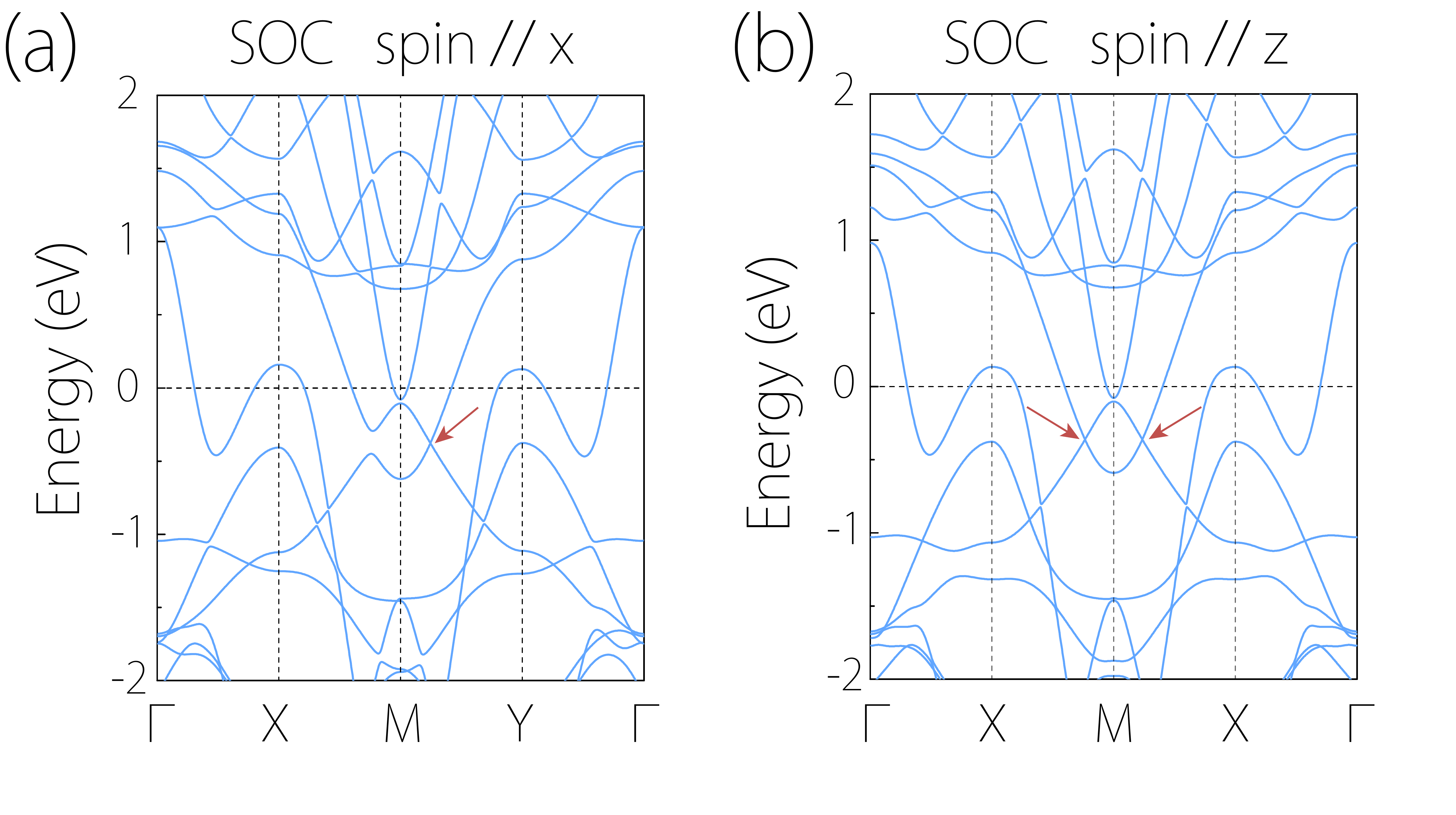}
	\caption{Band structures of $\gamma$-CoSe (with SOC included) with magnetic moments in the (a) $x$ and (b) $z$ direction, respectively. For better comparison, we repeat the path $X$-$M$ in (b). The arrow in (b) indicates the Weyl point. The arrows in (a) indicate the points on a Weyl loop.  }
	\label{fig_gamma-band-soc}
\end{figure}

\begin{figure}
	\includegraphics[width=0.42\textwidth]{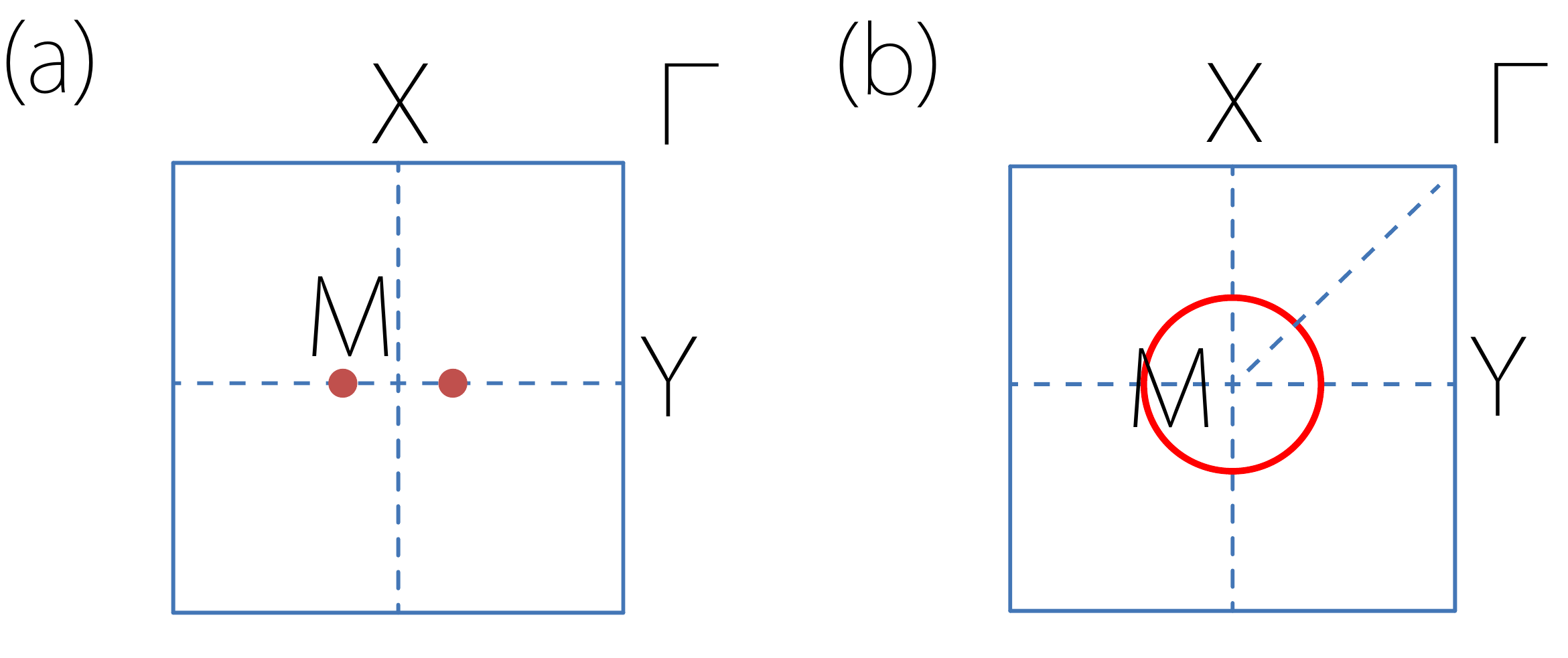}
	\caption{(a) Illustration of the location of the two magnetic Weyl points (red dots) in the ground state of $\gamma$-CoSe with moments along the $x$ direction. (b) The two Weyl points transform into a Weyl loop (red loop) when the moments are oriented along the $z$ direction. }
	\label{fig_gamma-loop}
\end{figure}

\subsection{$\gamma$-CoSe: Magnetic Weyl point versus Weyl loop}

$\gamma$-CoSe has a FM ground state with in-plane magnetization (along $x$). Figure~\ref{fig_gamma-band} shows its band structure in the absence of SOC. One observes that $\gamma$-CoSe is a FM metal. From the PDOS plot, the low-energy bands are dominated by the Co-$3d$ orbitals. Meanwhile, the Se-$4p$ orbitals also give a sizable contribution. The spin polarization at the Fermi level is about 42$\%$.

One notices that in Fig.~\ref{fig_gamma-band}(a), slightly below the Fermi level, the spin-minority bands form linear band crossings around the $M$ point. Scanning the band structure around this region shows the existence of a Weyl loop centered at $M$. As a Weyl loop, it is twofold degenerate. The loop is formed by an electron-like band and a hole-like band, hence it belongs to type-I loops according to its dispersion~\cite{li2017type}.

Like most previous 2D examples, the loop is destroyed when SOC is considered. However, as shown in Fig.~\ref{fig_gamma-band-soc}(a), although most points on the loop are gapped out, there remains a pair of Weyl points on $M$-$Y$. Note that distinct from Weyl points in 3D systems which are topologically stable, Weyl points in 2D must require extra symmetry protections. We find that the key symmetry here is the $C_{2x}$ symmetry, i.e., the twofold rotation along the magnetization direction. The two Weyl points are protected, as the two crossing bands have opposite $C_{2x}$ eigenvalues along $M$-$Y$. Stable Weyl points in 2D under SOC were only recently found in very few examples, such as in monolayer PtCl$_3$~\cite{you2019two} and in monolayer GaTeI~\cite{wu2019hourglass}. Our work offers another example, with the protecting symmetry different from the previous cases.

More interestingly, if we rotate the magnetization to the $z$ direction, the Weyl loop around $M$ can be recovered [see Fig.~\ref{fig_gamma-band-soc}(b)]. This is because in such a case, the system has a horizontal mirror $M_z$ preserved. The two crossing bands have opposite $M_z$ eigenvalues, hence the loop is protected even under SOC. This is a very interesting magneto-band-structure effect, namely, by controlling the magnetization direction, one can tune the transformation of topological band features, here between a pair of mangetic Weyl points and a magnetic Weyl loop, as illustrated in Fig.~\ref{fig_gamma-loop}.

\section{Discussion and CONCLUSION}

We have a few remarks before closing. First, in this work, we have discussed in detail three 2D CoSe structures, corresponding to the three with the lowest energies obtained from the extensive structural search. There exist other possible meta-stable structures with higher energies. (One of them, the $\delta$-CoSe, is discussed in~\cite{OtherPair}) Although the low energy ones have a better chance to be achieved, it is possible that, depending on the growth condition, the higher energy ones may also have an opportunity to be realized in experiment.

Second, in our study, we have adopted the commonly used $U$ value $U=4$ eV for Co $d$ orbitals~\cite{dalverny2010interplay}. In Sec.~\ref{beta}, we see that larger $U$ values can turn $\beta$-CoSe into an insulator. We have also check the effect of larger $U$ values (up to 6 eV) on $\alpha$- and $\gamma$-CoSe. We find that for $\gamma$-CoSe, the results remain qualitatively the same. As for $\alpha$-CoSe, very similar to the $\beta$ phase, larger $U$ values can turn the system into an AFM insulator. 

Third, as well developed methods for growing 2D materials,  the chemical vapor deposition and molecular beam epitaxy methods may be good choices for realizing the proposed 2D CoSe structures.
The monolayer $\alpha$-CoSe may also be obtained by the exfoliation method from the bulk tetragonal CoSe samples, which is commonly applied for making other 2D materials.

Fourth, we note that in previous experiments on tetragonal CoSe ultrathin films epitaxially grown on SrTiO$_3$(001) (STO) substrates~\cite{shen2018evolution,liu2018anti}, the samples are found to be paramagnetic, differing from our results here. A possible reason is that the properties of those CoSe samples could be affected by the STO substrate. The STO surface can form various reconstructed structures~\cite{Erdman2003} and it can strongly influence the 2D materials grown on it, as demonstrated in the prominent example of enhanced superconductivity of monolayer FeSe on STO~\cite{Wang2012}. Indeed, the strong coupling between CoSe films and STO substrate was inferred in experiment~\cite{shen2018evolution}.

Finally, we comment on a few experimental aspects for detecting our predicted effects. The band topology features of
the Weyl points and the Weyl loop in $\gamma$-CoSe can be directly imaged by the angle resolved photoemission spectroscopy (ARPES) technique. This technique has been successfully applied for many cases before, such as imaging the nodal loops in monolayer Cu$_2$Si~\cite{feng2017experimental} and CuSe~\cite{gao2018epitaxial}.  For {$\alpha$- and} $\beta$-CoSe, we have shown that strain can drive a phase transition from an AFM metal to an AFM insulator. For 2D materials, strain can be readily applied, e.g., by using a beam-bending apparatus~\cite{lee2008measurement} or by using an atomic-force microscope tip~\cite{conley2013bandgap}. For $\gamma$-CoSe, to control its magnetization direction, one can use an applied magnetic field or couple it to a magnetic substrate.

In conclusion, we have performed a comprehensive search for CoSe 2D structures. We have identified three lowest energy candidates and revealed their rich physical properties. We show that they possess different magnetic ground states. While $\gamma$-CoSe is FM,
$\alpha$- and $\beta$-CoSe are two rare examples of 2D AFM metals. We argue that this peculiar state is associated with the Fermi surface nesting features in their band structures.
$\alpha$-CoSe hosts a pair of anisotropic valleys in the valence band, which are connected by crystalline symmetry rather than $\mathcal{T}$. Hence it exhibits valley physics distinct from the well-known MoS$_2$-family materials, such as the pure strain control of the valley polarization. We also find that applied strain can drive rich metal-semiconductor and magnetic phase transitions in $\alpha$- and $\beta$-CoSe. $\gamma$-CoSe hosts a pair of magnetic Weyl points in its ground state, and a transformation from magnetic Weyl points to a magnetic Weyl loop can be induced by rotating the magnetization direction. Importantly, all these topological band features are robust against SOC. Our work uncovers a fascinating material platform for studying the interplay between magnetism, valley physics, correlation effects, and band topology in 2D. These predicted materials could have promising applications in electronics and spintronics.

\begin{acknowledgments}
The authors thank D. L. Deng for valuable discussions. This work is supported by the Singapore Ministry of Education AcRF Tier 2 (MOE2019-T2-1-001), National Natural Science Foundation of China (Grants No. 11504013 and No. 11974307), National Key Research and Development Program (No. 2019YFE0112000) and Zhejiang Provincial Natural Science Foundation (R21A040006, D19A040001). We acknowledge computational support from the Texas Advanced Computing Center and the National Supercomputing Centre Singapore.

B.T. and W.W. contributed equally to this work. 
\end{acknowledgments}

\bibliographystyle{apsrev4-1}{}
\bibliography{reference}

\end{document}